\documentclass[journal]{IEEEtran}
\usepackage[utf8]{inputenc}
\usepackage{balance}
\usepackage{booktabs}
\usepackage[T1]{fontenc}
\usepackage{physics,amsmath}
\usepackage{amsthm}
\usepackage{mathtools}
\usepackage{amsfonts}
\usepackage{amssymb}
\usepackage{graphicx}
\usepackage{makeidx}
\usepackage{csquotes}
\usepackage{cite}
\usepackage{xcolor}
\usepackage{subcaption}
\usepackage{float}
\usepackage{multicol}
\usepackage{multirow}
\usepackage{array}
\usepackage{pifont}
\usepackage{hyperref}
\usepackage[capitalise,nameinlink ,noabbrev]{cleveref}

\usepackage[utf8]{inputenc}
\usepackage[algo2e,ruled, vlined, linesnumbered]{algorithm2e}

\usepackage[justification=justified,singlelinecheck=false]{caption}
\usepackage{caption}
\usepackage[hyphenbreaks]{breakurl}
\begin{document}
	\title{\LARGE Enhancing Indoor and Outdoor THz Communications with Beyond Diagonal-IRS: Optimization and Performance Analysis}
\author{Asad Mahmood, Thang X. Vu,  Symeon Chatzinotas, Bj\"orn Ottersten \\Interdisciplinary Centre for Security, Reliability and Trust (SnT), University of Luxembourg\\
\{asad.mahmood, thang.vu, symeon.chatzinotas, bjorn.ottersten\}@uni.lu}%
 	\markboth{IEEE International Symposium on Personal, Indoor and Mobile Radio Communications 2024}%
 {Shell \MakeLowercase{\textit{et al.}}: Bare Demo of IEEEtran.cls for IEEE Journals} 
	\maketitle

\begin{abstract}
This work investigates the application of Beyond Diagonal Intelligent Reflective Surface (BD-IRS) to enhance THz downlink communication systems, operating in a hybrid: reflective and transmissive mode, to simultaneously provide services to indoor and outdoor users. We propose an optimization framework that jointly optimizes the beamforming vectors and phase shifts in the hybrid reflective/transmissive mode, aiming to maximize the system sum rate. To tackle the challenges in solving the joint design problem, we employ the conjugate gradient method and propose an iterative algorithm that successively optimizes the hybrid beamforming vectors and the phase shifts. Through comprehensive numerical simulations, our findings demonstrate a significant improvement in rate when compared to existing benchmark schemes, including time- and frequency-divided approaches, by approximately $30.5\%$ and $69.9\%$ respectively and even outperforms the STAR-IRS system by $76.99\%$. This underscores the significant influence of IRS elements on system performance relative to that of base station antennas, highlighting their pivotal role in advancing the communication system efficacy.
\end{abstract}


	\section{Introduction}
\IEEEPARstart{I}n the realm of future wireless communications, including beyond 5G (B5G) and 6G technologies, there is a drive to establish higher performance benchmarks and introduce novel application scenarios for societal digitization \cite{akyildiz2022terahertz,mahmood2022optimizing}. Terahertz (THz) communication has received considerable attention for its potential to deliver ultra-high data rates, thereby addressing the spectrum scarcity and capacity limitations faced by current communication systems\cite{zhu2022robust}. Despite its promise, the adoption of THz technology faces significant challenges due to the inherent properties of THz, such as severe attenuation by molecular absorption and limited penetration through wall/obstacle, which are characteristic of high-frequency propagation\cite{han2024thz}. To overcome these challenges,  intelligent reflecting surfaces (IRS) emerge as practical solution capable of adjusting their amplitude and phase to redirect signal transmission, thereby enhancing signal strength and coverage \cite{mahmood2023joint,8647620}.
\par
To fully exploit the advantages of IRS technology in THz communications, Hao et al. \cite{hao2021robust} focus on optimizing the weighted sum rate in THz multiple input multiple output (MIMO) systems through hybrid beamforming and IRS phase shifts. Zhao et al. \cite{zhao2023joint} proposed the time delay-based IRS scheme to mitigate the impact of beam quint in THz communication. Zhu et al. \cite{zhu2022robust} develop a robust beamforming strategy for IRS-assisted simultaneous wireless information and power transfer (SWIPT) in secure THz systems, with an emphasis on minimizing power consumption and adhering to outage constraints. Yuan et al. \cite{yuan2022secure} propose an approach for calculating the ergodic secrecy rate in THz-enabled IRS-assisted non-terrestrial network (NTN), considering atmospheric and phase challenges. Despite these advancements, all of these research efforts consider the traditional IRS architecture, which is capable of transmitting signals in one direction. Moreover, as per \cite{huawei2024,cisco2024}, a considerable amount of mobile data traffic, estimated to be between $80\%$ and $96\%$, is generated indoors due to the prevalent indoor lifestyle of users. This underscores the need for efficient outdoor-to-indoor communication solutions.
\par
To address this challenge, Xing et al. \cite{xing2021millimeter} and Chen et al. \cite{chen2023channel} have propelled the development of efficient architecture focusing on significant advancements in channel modeling and measurement for indoor THz and sub-THz communications. Concurrently, Yildirim et al. \cite{yildirim2020modeling} and Shaikh et al. \cite{shaikh2023energy} have made notable contributions to this domain, primarily concentrating on the deployment of multiple IRS units, which, however, may not align with the principles of resource efficiency. To overcome the limitations of traditional IRS, Beyond Diagonal Intelligent Reflective Surfaces (BD-IRS) are introduced by Li et al. \cite{li2023beyond, li2022beyond1}, Nerini et al. \cite{nerini2023discrete}, and Soleymani et al. \cite{soleymani2023optimization} as a novel advancement and demonstrate superiority over traditional IRS. Meanwhile, the unique demands of THz networks, particularly the issue of higher frequency attenuation within buildings, require thorough consideration and investigation, as done by Wang et al. \cite{wang2023simultaneously} on simultaneous transmission and reflecting (STAR)-IRS for enabling THz communication across varied user groups. This joint effort underscores a strong motivation to advance communication technologies, highlighting BD-IRS as a promising solution for the future's communication system, like THz.
\par
Hence, motivated by the preceding discussion, this study investigates the utilization of fully connected BD-IRS operating in a hybrid mode, incorporating both reflective and transmission functionalities. This configuration enables the enhanced simultaneous servicing of indoor and outdoor users by efficiently reflecting and transmitting signals, thereby optimizing resource utilization and improving overall network performance. Subsequently, the main contributions of this work are summarized as follows:
\begin{itemize}
\item We develop a joint optimization framework aimed at facilitating simultaneous service to both indoor and outdoor users. This is achieved by jointly optimizing the hybrid beamforming vectors at the THz base station (BS) and the phase shifts of the BD-IRS operating in the hybrid mode.
\item To address the optimization challenge, we decompose the joint problem into subproblems, focusing on hybrid beamforming and IRS phase-shift control. These are tackled iteratively using the Block Coordinate Descent (BCD) method, allowing for a systematic solution to the original complex optimization problem.
\item Our results indicate that the proposed approach significantly outperforms existing benchmark schemes, with improvements in rate of approximately 30.50\% and 69.9\%, over the time- and frequency-division approaches and 76.99\% over the STAR-IRS, thereby highlighting the critical impact of IRS elements on system performance compared to BS antennas.
\end{itemize}
The paper is structured as follows. Section \ref{SM} details the system model and problem formulation. Section \ref{PSol} outlines the proposed solution. The numerical results are presented in Section \ref{RnD}, followed by the conclusions in Section \ref{Con}.
\section{System Model}\label{SM}
\begin{figure}
\centering
\includegraphics[width=2.5in]{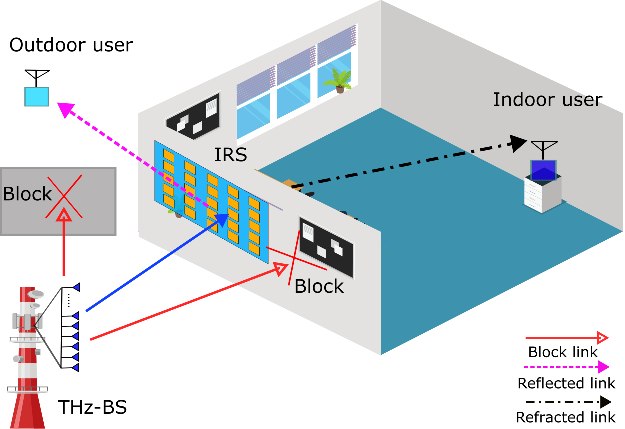}
	\caption{System Model}
	\label{fig:SM}
\end{figure}

This study explores the utilization of BD-IRS in fully connected architecture to enhance THz downlink communication systems to provide service to $N$ single antenna indoor / outdoor users, denoted $\mathcal{N}=\{1,2,\ldots, N\}$, as depicted in Fig. \ref{fig:SM}. The THz base station (BS) is equipped with $M$ antennas and $M_{\text{RF}}$ RF chains where $M_{RF}\!\! \leq\! M$. It is noteworthy that the number of users served is inherently limited by the number of RF chains. BD-IRS comprises $K$ elements and denoted by $\mathcal{K}=\{1,2,\ldots, K\}$. Users are divided into two distinct groups based on their physical location: reflective (outdoor) users, indexed by $\mathcal{N}_r\subseteq \mathcal{N}$, and transmissive (indoor) users, indexed by $\mathcal{N}_t\subseteq \mathcal{N}$. Under the assumption of the acquisition of perfect channel state information (CSI) achieved by channel estimation techniques \cite{zhao2023joint,nguyen2022channel}, this scenario posits an upper bound solution. Since the number of RF chains is less than the number of antennas, a hybrid beamforming design must be adopted that entails initial signal processing via a digital beamformer $\mathbf{V}^{\text{BB}} \in \mathbb{C}^{M_{\text{RF}}\times N}$, followed by an analog beamformer $\mathbf{V}^{\text{RF}} \in \mathbb{C}^{M\times M_{\text{RF}}}$. Furthermore, due to the blockage, the BS can only transmit data to the users via the BD-IRS, where it concurrently reflects and transmits the signal to the intended user. In  this context, the BD-IRS scattering matrix can be denoted as $\boldsymbol{\Theta}_{i_n} \in \mathbb{C}^{K \times K}$, where $i_n = t$ if $n \in \mathcal{N}_t$ and $i_n = r$ if $n \in \mathcal{N}_r$ subject to the constraint $\boldsymbol{\Theta}_{r}^H \boldsymbol{\Theta}_{r} + \boldsymbol{\Theta}_{t}^H \boldsymbol{\Theta}_{t} = \mathbf{I}_{K}$, which ensures energy conservation in both reflection and transmission processes. The signal received from the $n$-th user is given as follows:
\begin{equation}
\label{Eq_RS}
y_n = \boldsymbol{h}_n^H \boldsymbol{\Theta}_{i_n} \boldsymbol{G} \boldsymbol{w}_n x_n\! + \!\sum_{n'\ne n} \boldsymbol{h}_n^H \boldsymbol{\Theta}_{i_{n}} \boldsymbol{G} \boldsymbol{w}_{n'} x_{n'} + \eta_n.
\end{equation}

 In \eqref{Eq_RS}, the beamforming vector $\boldsymbol{w}_n = \mathbf{V}^{\text{RF}} \mathbf{V}^{\text{BB}}_n \in \mathbb{C}^{M\times 1}$ represents the composite beamforming effect for the $n$-th user. The channel vectors $\boldsymbol{h}_n \in \mathbb{C}^{K\times 1}$ and $\boldsymbol{G} \in \mathbb{C}^{K\times M}$ represent the channel from the IRS to the user and from the BS to the IRS, respectively.
The transmitted signal for the $n$-th user, $x_n$, adheres to the condition $\mathbb{E}\{x_nx_n^*\}=1$, and the additive white Gaussian noise at the user is denoted by $\eta_n$. Given the substantially lower power gain of non-line-of-sight (NLoS) paths compared to line-of-sight (LoS) paths in THz communications, this analysis primarily concentrates on the LoS channel model\cite{zhu2022robust}. Hence, the channel $\boldsymbol{G}$ is modeled as $\boldsymbol{G} = q(f_c,d_1)\Bar{\boldsymbol{G}}$, where $q(f_c,d_1) = \frac{c}{4\pi f_c d_1}e^{-\frac{1}{2}\tau(f_c)d_1}$ encapsulates the effects of free-space path loss and medium absorption. Here, $c$ signifies the speed of light, $f_c$ is the carrier frequency, $d_1$ represents the distance from the BS to the IRS, and $\tau(f_c)$ is the medium's absorption coefficient obtained from the high-resolution transmission (HITRAN) database \cite{rothman2009hitran}. The antenna array response vectors at the transmitter and IRS are represented by $\boldsymbol{\Bar{G}} =  \boldsymbol{a}_{\text{tx}}(\vartheta_{\text{tx}})\boldsymbol{a}^H_{\text{rx}}(\vartheta_{\text{rx}})$, defined for the respective angles of arrival and departure.
\begin{align}
\!\!\!\!\!    \boldsymbol{a}_{rx}(\vartheta_{rx}) \!&=\! \!\left[1, e^{j \pi \vartheta_{rx}}, e^{j 2 \pi \vartheta_{rx}}, \ldots, e^{j(M-1) \pi \vartheta_{rx}}\right]^T, \\
\!\!\!\!\!     \boldsymbol{a}_{tx}(\vartheta_{tx}) &= \left[1, e^{j \pi \vartheta_{tx}}, e^{j 2 \pi \vartheta_{tx}}, \ldots, e^{j(K-1) \pi \vartheta_{tx}}\right]^T.
\end{align}
Here, $\vartheta_i \!= \!\!2 d_0 f_c \sin(\phi_i) / c, i \in \{{rx}, tx\}$, with $d_0$ representing antenna spacing, and $\phi_i \in [-\pi / 2, \pi / 2]$ denoting angle of departure (AoD) and angle of arrival (AoA), respectively. The channel vector $\boldsymbol{h}_n= \!q(f_c,d_2)\boldsymbol{a}_{\text{tx}}(\vartheta_{\text{tx}})$. Here $d_2$ denotes the distance between the IRS and users.  Moreover, the signal-to-interference plus noise ratio (SINR) of $n-$th user can be expressed as: 
\begin{equation}
\label{SINR}
\gamma_n=\frac{| \boldsymbol{h}_n^H \boldsymbol{\Theta}_{i_{n}}\boldsymbol{G} \boldsymbol{w}_n|^2}{{\sum}_{{n'}\ne n} |\boldsymbol{h}_n^H \boldsymbol{\Theta}_{i_{n}} \boldsymbol{G} \boldsymbol{w}_{n'}|^2+\sigma^2}.
\end{equation}
\subsection{Problem Formulation}
In this work, we aim to maximize the sum rate of the users by jointly optimizing the hybrid beamforming vector $\{\mathbf{V}^{\text{RF}},\mathbf{V}^{\text{BB}}\}$, followed by an IRS scattering matrix $\boldsymbol{\Theta}_{t,r}$. Formulating the joint optimization problem as:
\begin{subequations}
\label{JP}
\begin{align}
\max_{\mathbf{V}^{\text{RF}},\mathbf{V}^{\text{BB}},\boldsymbol{\Theta}_r,\boldsymbol{\Theta}_t}&  {\sum}_{n=1}^{N} \log_2\left(1+\gamma_n\right) \label{OF}\\
\text{s.t.} & \boldsymbol{\Theta}_{r}^H \boldsymbol{\Theta}_{r} + \boldsymbol{\Theta}_{t}^H \boldsymbol{\Theta}_{t} = \boldsymbol{I}_{{K}}, \label{C1}\\
& \left\|\mathbf{V}^{\mathrm{RF}} \mathbf{V}^{\mathrm{BB}}\right\|_F^2 \leq P_{\max},\label{C2}  \\
& \left|\mathbf{V}^{\mathrm{RF}}(i, j)\right|=1, \forall i, j\label{C3},
\end{align}
\end{subequations}
where $\gamma_n$ is given in \eqref{SINR},  $P_{\max}$ denotes the maximum allowable transmit power at the BS. The \eqref{JP} poses substantial challenges because of nonconvex constraints along with the unitary nature of the IRS and the nonlinearity of the objective function, as well as the coupling of variables detailed in \eqref{C2}. To overcome these, we reformulate \eqref{JP} into a more manageable form by using fractional programming techniques. Following this, an iterative method is employed to solve the reformulated problem efficiently. This approach facilitates a systematic enhancement of the sum rate for the users, achieved through the meticulous optimization of both the hybrid beamforming vector and the IRS scattering matrix.

\section{Proposed Solution}

 \label{PSol}
This section presents a proposed framework, in which we initially transform the objective function to more tractable by introducing auxiliary variables $\boldsymbol{\beta} \in \mathbb{R}^N = [\beta_1, \beta_2, \ldots, \beta_N]^T$ and $\boldsymbol{\alpha} \in \mathbb{C}^N = [\alpha_1, \alpha_2, \ldots, \alpha_N]^T$. Employing a quadratic transformation to convert the fractional SINR terms into an integer expression, we reformulate the  \eqref{OF} as follows: 
\cite[Sec. IV-C]{shen2018fractional}
\begin{equation}
\label{OFC}
\mathcal{F}(\boldsymbol{\mathcal{X}}) = \sum_{n=1}^N (\log_2(1 + \beta_n) - \beta_n + \Gamma_n - \Xi_n),
\end{equation}
where ${\small \boldsymbol{\mathcal{X}} = \{\mathbf{V}^{\text{RF}},\mathbf{V}^{\text{BB}}, \boldsymbol{\Theta}_r, \boldsymbol{\Theta}_t, \boldsymbol{\beta}, \boldsymbol{\alpha}\}}$ is the shorthand notation of the variables, ${\small \Gamma_n \!=\! 2\sqrt{1 + \beta_n} \Re{\alpha_n^\dagger \bar{\boldsymbol{h}}_n^H \boldsymbol{w}_n}}$, and ${\small \Xi_n \!= |\alpha_n|^2 \sum_{n' \!\ne n} |\bar{\boldsymbol{h}}_n^H \boldsymbol{w}_{n'}|^2 \!+ \!\sigma^2}$; ${\small \bar{\boldsymbol{h}}_n = (\boldsymbol{h}_n^H \boldsymbol{\Theta}_{i_n} \boldsymbol{G})^H}$. Moreover, the new optimization problem expressed as:
	\begin{subequations}
		\label{XXX}
		\begin{align}
			{\max}_{\boldsymbol{\mathcal{X}}}  \quad \mathcal{F}(\boldsymbol{\mathcal{X}})  \;\;
			\text{s.t.}  \quad \text{\cref{C1,C2,C3}}.
		\end{align}
	\end{subequations}
Considering the multivariate complexity of \eqref{XXX}, we adopt BCD to separate it into distinct subproblems. These subproblems are then iteratively addressed, with detailed explanations provided in the subsequent subsections.
	
	\subsection{Auxiliary Parameter Optimization}
	\label{UAPO}
	Under given $\mathbf{V}^{\text{RF}},\mathbf{V}^{\text{BB}}$, $\boldsymbol{\Theta}_r$, and $\boldsymbol{\Theta}_t$, we  optimize $\boldsymbol{\beta}$ and $\boldsymbol{\alpha}$ in the convex framework of \eqref{XXX}. Setting the derivatives $\frac{\partial\mathcal{F}}{\partial\beta}=0$ and $\frac{\partial\mathcal{F}}{\partial\alpha}=0$ yields:
 \begin{equation}
      \beta^*_n = \gamma_n,\; \alpha^*_n = \frac{\sqrt{1+\beta_n}\bar{\boldsymbol{h}}_n^H \boldsymbol{w}_n}{\sum_{n' \ne n} |\bar{\boldsymbol{h}}_n^H \boldsymbol{w}_{n'}|^2 + \sigma^2}.
 \end{equation}
	\subsection{Hybrid Beamforming}
	In the realm of hybrid beamforming, the convexity of \eqref{OFC} facilitates the formulation of the optimization problem for $\mathbf{V}^{\text{RF}}$ and $\mathbf{V}^{\text{BB}}$ as follows:
	\begin{subequations}
		\label{JP2}
		\begin{align}
			\!\!\!\!\!\max_{\mathbf{V}^{\text{RF}},\mathbf{V}^{\text{BB}}} \mathcal{F}(\mathbf{V}^{\text{RF}},\mathbf{V}^{\text{BB}}) \;\;
			\text{s.t.}   \text{\cref{C2,C3}}.
		\end{align}
	\end{subequations}
Despite the objective function's convexity, the coupling of decision variables presents a challenge to deriving optimal solutions. To mitigate this, subproblems are formulated, with a subset of variables being held constant as each subproblem is solved iteratively.
Hence, the analog beamforming subproblem is specified as follows:
	\begin{subequations}
		\label{JP2a}
		\begin{align}
			&\max_{\mathbf{V}^{\text{RF}}} \quad \mathcal{F}(\mathbf{V}^{\text{RF}}) \quad \;\;
			\text{s.t.}  \; \text{\cref{C2,C3}}.
		\end{align}
	\end{subequations}
Similarly, the subproblem for digital beamforming is outlined as:
	\begin{subequations}
		\label{JP2b}
		\begin{align}
			&\max_{\mathbf{V}^{\text{BB}}} \quad \mathcal{F}(\mathbf{V}^{\text{BB}}) \;\;
			\text{s.t.}  \quad \text{\cref{C2}}.
		\end{align}
	\end{subequations}
Subproblems \eqref{JP2a} and \eqref{JP2b} are convex and can be solved iteratively using standard optimization tools, e.g., CVX.
\subsection{IRS Phase Shift Optimization}
Given $\mathbf{V}^{\text{RF}},\mathbf{V}^{\text{BB}}$, $\boldsymbol{\beta}$, and $\boldsymbol{\alpha}$, we optimize $\boldsymbol{\Theta}_r$ and $\boldsymbol{\Theta}_t$ as follows:
\begin{subequations}
 \small
\begin{align}
\!\!\!\max_{\boldsymbol{\Theta}_r,\boldsymbol{\Theta}_t} & \!\sum_{{i_n}}\! \left( 2\Re\{\text{Tr}(\boldsymbol{\Theta}_{i_n} \!\boldsymbol{X}_{i_n})\}\! -\! \text{Tr}(\boldsymbol{\Theta}_{i_n} \boldsymbol{Y} \boldsymbol{\Theta}_{i_n}^H \boldsymbol{Z}_{i_n})\! \right), \\
\text{s.t.} & \quad \text{\cref{C1}}, \notag
\end{align}
\end{subequations}
where $\boldsymbol{X}_{i_n}\in \mathbb{C}^{K\times K}=\sum_{n\in \mathcal{N}_{i_n}}\sqrt{1+\beta_n}\alpha_n\boldsymbol{G}\mathbf{V}^{\text{RF}} \mathbf{V}^{\text{BB}}_n\boldsymbol{h}_n^H$, $\boldsymbol{Y}\in \mathbb{C}^{K\times K}=\sum_{n'\ne n}\boldsymbol{G}\mathbf{V}^{\text{RF}} \mathbf{V}^{\text{BB}}_{n'}(\boldsymbol{G}\mathbf{V}^{\text{RF}} \mathbf{V}^{\text{BB}}_{n'})^H$, and $\boldsymbol{Z}_{i_n}\in \mathbb{C}^{K\times K}=\sum_{n\in \mathcal{N}_{i_n}}|\alpha_n|^2\boldsymbol{h}_n\boldsymbol{h}_n^H$.  
Building on \cite[Sec. IV-E]{li2022beyond1}, this simplify to $\boldsymbol{\Theta} = [\boldsymbol{\Theta}^H_{t}, \boldsymbol{\Theta}^H_{r}]^H$, $\boldsymbol{X} = [\boldsymbol{X}_{t}, \boldsymbol{X}_{r}]$, and $\boldsymbol{Z} = \text{blkdiag}(\boldsymbol{Z}_{t}, \boldsymbol{Z}_{r})$, leading to the subproblem for $\boldsymbol{\Theta}$ as:
	\begin{subequations}
		\begin{align}
			\max_{\boldsymbol{\Theta}}\!\!\!\!\! & \quad\mathcal{F}(\boldsymbol{\Theta})\!= 2\Re{\text{Tr}(\boldsymbol{\Theta}{\boldsymbol{X}})}\!-\!\text{Tr}(\boldsymbol{\Theta}\boldsymbol{Y}\boldsymbol{\Theta}^H\boldsymbol{Z}),\label{FOF} \\
			\text{s.t.} & \quad \boldsymbol{\Theta}^H\boldsymbol{\Theta} = \boldsymbol{I}_{{K}}.
		\end{align}
	\end{subequations}
Utilizing the conjugate gradient ascent method \cite{Abr08_PhD_thesis}, this strategy leverages the IRS elements to streamline the optimization of phase shifts. The initial step involves computing the Euclidean gradient of \eqref{FOF}, expressed as $\nabla\mathcal{F}(\boldsymbol{\Theta}) = 2\boldsymbol{X}^H  -  2\boldsymbol{Z}\boldsymbol{\Theta}\boldsymbol{Y}$. This computation aids in determining the Riemannian gradient at $\boldsymbol{\Theta}$, formulated as: $\boldsymbol{J}(\boldsymbol{\Theta}) = \nabla\mathcal{F}(\boldsymbol{\Theta})\boldsymbol{\Theta}^H - \boldsymbol{\Theta}\nabla\mathcal{F}(\boldsymbol{\Theta})^H$. The Riemannian gradient, $\boldsymbol{J}(\boldsymbol{\Theta})$, then facilitates the calculation of the rotation matrix $\boldsymbol{R} = \boldsymbol{I} + \mu\boldsymbol{J} + \frac{\mu^2}{2}\boldsymbol{J}^2$, with $\mu$ serving as the control parameter for convergence. This rotation matrix is subsequently employed to iteratively update the IRS phase shift matrix as $\boldsymbol{\Theta}_l^{j+1} = \boldsymbol{J}(\boldsymbol{\Theta}^{j})\boldsymbol{\Theta}^{j}$.	
Algorithm \ref{Algo1} outlines its operations and sets the worst case per iteration complexity at $\mathcal{O}(N^2K^2 + I_1(M^{3.5} + (M_{\text{RF}}N)^3) + I_2K^3)$, with $I_1$ and $I_2$ as the maximum iterations for hybrid beamforming and the conjugate gradient method, respectively. The function $\mathcal{F}(\boldsymbol{\mathcal{X}}^{t+1}) \geq \mathcal{F}(\boldsymbol{\mathcal{X}}^t)$ demonstrates that $\mathcal{F}$ iteratively converges towards a local optimum, with the pursuit of the global maximum identified as an area for further research.

	\begin{algorithm2e}
		\small
		\SetAlgoLined
		\textbf{Initialization:} Initialize function value $\mathcal{F}_o=0$, IRS phase shift matrices $\boldsymbol{\Theta}^{j}=\mathbf{I}$, control parameters $\boldsymbol{\alpha}$, $\boldsymbol{\beta}$, $\boldsymbol{W}$, and step size $\mu=1$.\\
		\While{$|\mathcal{F}_i-\mathcal{F}_{i+1}|\le \epsilon$}{
			Update $[\mathbf{V}_i^{\text{RF}},\mathbf{V}_i^{\text{BB}}]$ by solving \eqref{JP2}.\\
				\Repeat{convergence}{
					Compute Euclidean space gradient: $\nabla\mathcal{F}=\frac{\partial \mathcal{F}}{\partial \boldsymbol{\Theta}^*}(\boldsymbol{\Theta}^j)$.\\
					Determine Riemannian space gradient direction: $\boldsymbol{J}(\boldsymbol{\Theta}^j) = \nabla\mathcal{F}(\boldsymbol{\Theta}^j){\boldsymbol{\Theta}^j}^H - \boldsymbol{\Theta}^j\nabla\mathcal{F}(\boldsymbol{\Theta}^j)^H$.\\
					\If{$\|\boldsymbol{J}(\boldsymbol{\Theta}^j)\|_{\boldsymbol{\Theta}^j}^2$ is sufficiently small}{
						\textbf{break}.
					}
					Compute rotation matrices: $\boldsymbol{R}^j=\boldsymbol{I}+\mu \boldsymbol{J}^j+(\mu \boldsymbol{J}^j)^2 / 2+(\mu \boldsymbol{J}^j)^3 / 6$, $\boldsymbol{U}^j=\boldsymbol{R}^j \boldsymbol{R}^j$.\\
					Adjust $\mu$ to satisfy gradient reduction conditions using $\boldsymbol{R}^j$ and $\boldsymbol{U}^j$.\\
					Update $\boldsymbol{\Theta}^{j+1} = \boldsymbol{J}(\boldsymbol{\Theta}^{j})\boldsymbol{\Theta}^{j}$, increment $j$.
				}
				Store optimal value of $\boldsymbol{\Theta}$.
			Update $\mathcal{F}_i$ by solving \eqref{OF} and adjust auxiliary parameters as per section \ref{UAPO}.
		}
		\caption{\textbf{{Iterative \!Algorithm to\! solve\!} \eqref{JP}}}
		\label{Algo1}
	\end{algorithm2e}
 \begin{table}
    \centering
    \caption{Simulation Parameters}
    \label{tab:SP}
    \begin{tabular}{@{}llll@{}}
        \hline
        Parameter & Value & Parameter & Value \\
        \hline
        Area Length & $100m^2$ meters & N & 4 \\
        $f_c$ & $[0.1-1]$Thz & B & 1MHz \\
        $N_o$ & -174 dBm/Hz& K&[25-100]  \\
        $M$ & [49-225]& $P_{max}$&[15-30] dBm  \\
        \hline
    \end{tabular}
\end{table}
\begin{figure}
	\centering
	\includegraphics[width=0.74\linewidth]{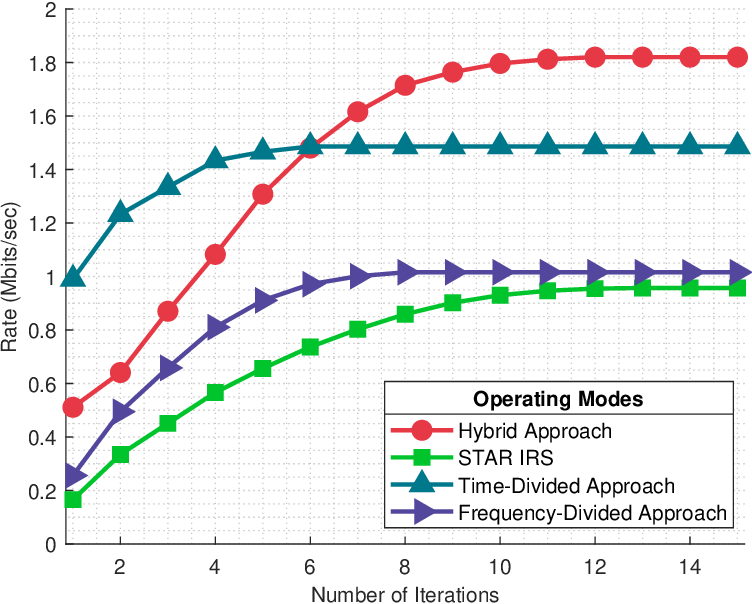}
	\caption{Convergence Analysis}
	\label{fig:R1}
\end{figure}

\begin{figure}
	\centering
	\includegraphics[width=0.74\linewidth]{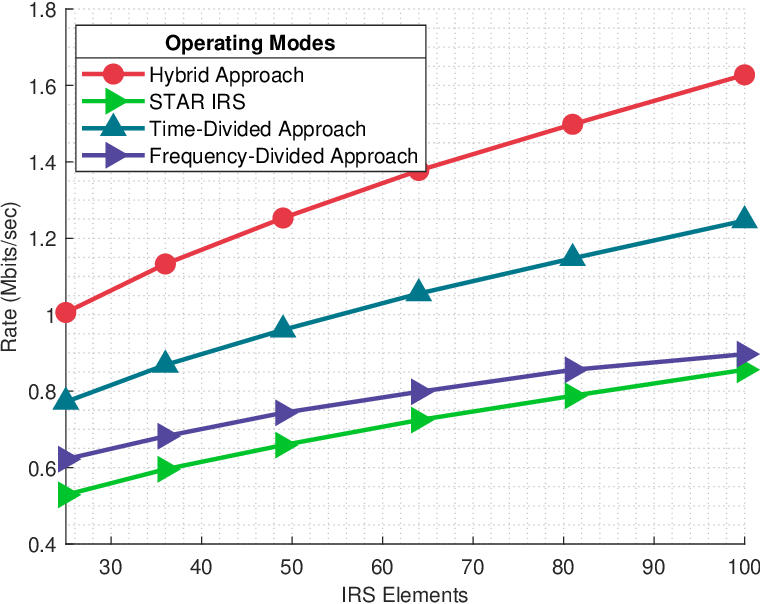}
	\caption{Operating Mode Vs IRS Elements}
	\label{fig:R2}

\end{figure}

\begin{figure}
	\centering
	\includegraphics[width=0.74\linewidth]{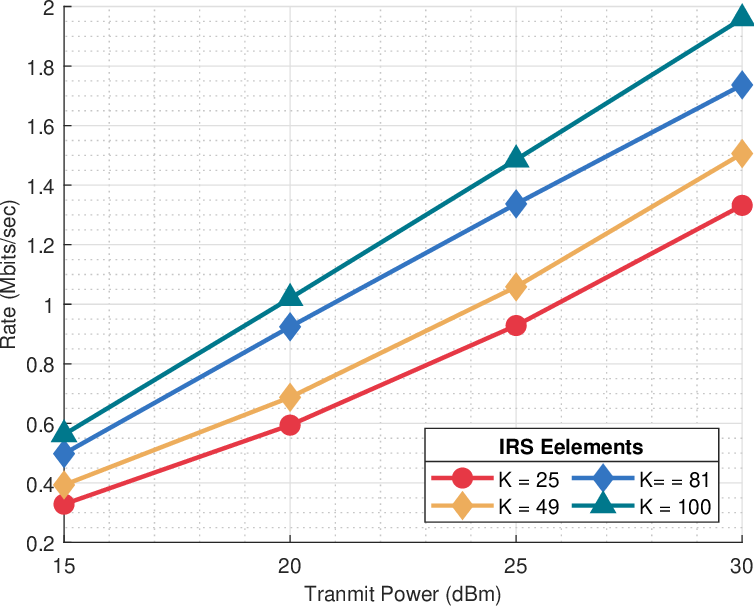}
	\caption{Transmit Power Vs IRS Elements}
	\label{fig:R3}

\end{figure}
\begin{figure}
	\centering
	\includegraphics[width=0.74\linewidth]{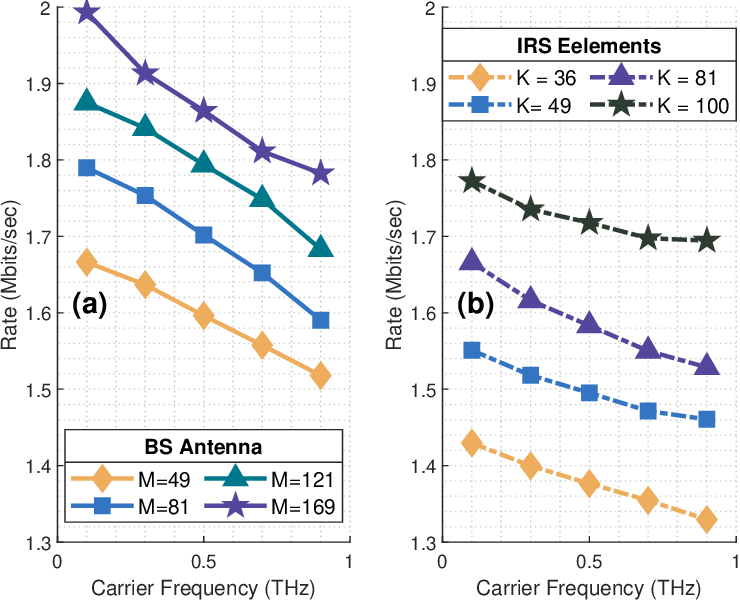}
	\caption{Carrier Frequency Vs BS Antennas}
	\label{fig:R4}
\end{figure}

\section{Performance Evaluation}
\label{RnD}
In this section, the performance of a fully connected BD-IRS to enhance THz communication systems is evaluated. We compare the hybrid transmissive/reflective design with two reference schemes: time-divided (TDMA) and frequency-divided (FDMA). In the former,  IRS consecutively switches between transmissive and reflective mode to serve all the users. In the latter, two user groups are simultaneously served in two orthogonal frequency bandwidths. The results are also compared with those of the STAR IRS \cite{wang2023simultaneously}. Furthermore, the analysis examines the impact of various simulation parameters, including the IRS performance across different frequency bands. Simulation parameters are listed in Table \ref{tab:SP}.
\par
The proposed scheme's effectiveness is determined by the algorithm's convergence. Through extensive simulation, average results are generated and plotted across different operating modes over iterations. Fig. \ref{fig:R1} shows that, as the algorithm iterates, it converges to a stable point by setting $N_r=N_t$ and $M=K$. Notably, the proposed scheme outperforms others due to limitations in reflective and transmissive IRS configurations, respectively. Time-divided allows reflective and transmissive users to receive signals at different intervals, whereas frequency-division serves both user groups concurrently but at different bandwidths. In contrast, the hybrid approach allows for simultaneous service to both user sets, resulting in a higher average rate than alternatives.
\par
The comparative analysis, illustrated in Fig. \ref{fig:R2} under consistent settings, evaluates the efficacy of the hybrid approach against time-divided, frequency-divided strategies and STAR IRS across a range of IRS element counts. The findings underscore the distinct advantage of the fully connected hybrid approach, which surpasses the time-divided, frequency-divided, and STAR IRS methods by approximately 30.50\% , 69.9\% and 76.99\%, respectively, demonstrating its notable superiority.
\par
Fig. \ref{fig:R3} illustrates the influence of transmit power and IRS elements on the system's performance. The results reveal that both factors contribute to enhancing system performance. Notably, increasing power levels exhibit a more pronounced effect on system performance compared to escalating the number of IRS elements. Specifically, power levels exhibit an average percentage increase of $55.98\%$, while the increase attributable to IRS elements is approximately $11.94\%$.
\par
Fig. \ref{fig:R4} showcases the effects of carrier frequency and the number of BS antennas or IRS elements on system performance. The findings in \ref{fig:R4}a indicate that an increase in carrier frequency leads to higher absorption path loss, resulting in decreased performance. On average, an increase in carrier frequency leads to a system performance change of about $-2.44\%$, whereas increasing the number of antennas results in an average performance improvement of approximately $5.44\%$. This underscores the beneficial role of additional antennas in enhancing system performance. Similarly, \ref{fig:R4}(b) presents consistent trends with varying numbers of IRS elements. Here, carrier frequency changes bring about a performance change of roughly $-1.57\%$, and increasing the number of IRS elements leads to an average performance improvement of approximately $7.44\%$ demonstrating its effectiveness as compared to BS antennas.

\section{Conclusion}
\label{Con}
In conclusion, this research delineates the efficacy of implementing fully connected BD-IRS in a hybrid operational mode to enhance THz downlink communications and provide simultaneous service to both indoor and outdoor single-antenna users. For this, we formulate the joint optimization problem for hybrid beamforming at the THz BS along with the BD-IRS phase shifts to maximize the sum rate. Leveraging the conjugate gradient method, the research meticulously navigates the complexities of the optimization landscape, presenting a coherent strategy for addressing these challenges. Through rigorous numerical simulations, the results affirm the proposed method's substantial enhancement in system performance compared to traditional benchmarks, including time- and frequency-divided approaches and STAR-IRS, by demonstrating significant improvements in terms of rate by approximately 30.50\% , 69.9\% and 76.99\%, respectively. This conclusive evidence underlines the proposed approach's efficacy and establishes a robust foundation for future advancements in THz communication networks. Future work may explore imperfect CSI and hardware limitations while enhancing energy efficiency in multi-user scenarios.

\section{Acknowledgement}

This work is funded by the Luxembourg National Research Fund (FNR) as part of the CORE program under project RISOTTI C20/IS/14773976.

\bibliographystyle{IEEEtran}
\bibliography{ReferenceBibFile}

\end{document}